\documentstyle[11pt,newpasp,twoside,epsf]{article}

\begin{document}

\title{Weak Lensing Observations of High-Redshift Clusters of Galaxies}
\author{D. Clowe}
\affil{Max-Planck-Institut f\"ur Astrophysik,
Karl-Schwarzschild-Str. 1, D-85748 Garching, Germany}
\author{G. Luppino, N. Kaiser, and I. Gioia}
\affil{Institute for Astronomy, University of Hawaii, 2680 Woodlawn Drive,
Honolulu, HI 96822}

\begin{abstract}
We present results of a weak gravitational lensing survey of six X-ray
selected high-redshift clusters of galaxies.  We find that the masses
of the clusters derived from weak lensing are comparable to those
derived from the X-ray observations.  We show that many of the
clusters have significant substructure not observed in the X-ray
observations and that for the more massive clusters a singular
isothermal sphere does not provide a good fit to the radial mass profile.
\end{abstract}

\section{Introduction}
High-redshift clusters of galaxies are very powerful tools for
testing cosmological and structure formation models.  Both the expected
number density of massive clusters at $z>0.5$ and the amount of
substructure contained within the clusters depend very strongly on
the mass density of the universe (Bahcall, Fan, \&\ Cen 1997).  
X-ray observations have proven to be a very efficient means to detect
these clusters.  Unlike optical surveys, in which one can find an
overdensity which is a superposition of unrelated galaxies, clusters
are detected in X-rays emitted from gas heated during infall into a large
potential well.  While masses derived from the X-ray observations have
been used to apply constraints to cosmological models (Henry 1997),
they are subject to an uncertainty in that they, like dynamical
mass estimates from cluster galaxy redshifts, depend on an assumed
dynamical state of the cluster (Evrard, Metzler, \&\ Navarro 1996).
Weak gravitational lensing, however, has no such dependence, and thus can,
in theory, provide mass estimates independent of any assumptions
regarding the cluster lens.  

We, therefore, have undertaken an optical survey of high-redshift,
X-ray selected clusters of galaxies to perform a weak lensing analysis
on the clusters.  Our primary goals are to measure the masses of the
clusters without assumptions regarding the degree of virialization of the
clusters and to detect any substructure in the clusters which would be
indicative of the clusters still undergoing initial formation.  We
have, to date, observed six $z>0.5$ clusters, five from the Einstein Medium
Sensitivity Survey (Gioia \&\ Luppino 1994) and one from the ROSAT
North Ecliptic Pole survey (Henry {\it et al.} 1998).  All six
clusters have deep ($> 1.5$ hours) exposures in $R$-band from the Keck
Observatory as well as shallower $I$- and $B$-band images from the
UH88$^{\prime\prime}$ telescope.

\section{Weak Lensing Analysis}

A weak lensing signal is detected in a field by measuring the
ellipticities of background galaxies and looking for a statistical
deviation from an isotropic ellipticity distribution.  We used a
hierarchical peak finding algorithm on the $R$-band images to detect
faint galaxies and measure their magnitudes and second moments of the
flux distribution.  The second moments were used to calculate
ellipticities for each object, which were then corrected for psf
anisotropies and reduction of the ellipticity due to psf smearing
using techniques originally developed in Kaiser, Squires, \&\
Broadhurst (1993).  Details of the data
reduction process can be found in Clowe {\it et al.} (1999) and a
review of these and other techniques can be found in Mellier (1999).

Once the ellipticities are corrected for psf smearing 
they can be used to measure the shear caused by the
gravitational lensing exerted by the cluster.  The shear field is
quite noisy because of the intrinsic ellipticity distribution of the
background galaxies, which is the dominant source of random error in
weak lensing analysis.  The shear can then be converted to the
convergence, $\kappa$, which is the surface mass density of the lens
divided by $\Sigma _{crit}$, where
\begin{equation}
\Sigma _{crit} = {c^2 \over 4 \pi  G}{D_{os} \over D_{ol} D_{ls}}.
\end{equation}
The $D$'s are angular distances between the observer, the lens (cluster), and
the source (background galaxy).  For low redshift clusters ($z<0.3$),
$\Sigma_{crit}$ is effectively constant for $z_{bg} > .8$, where most
background galaxies are located.  For high redshift clusters,
however, one must know the redshift distribution of the background
galaxies to convert the convergence to a surface density.  As the
redshift distribution of the galaxies used in our sample ($23<R<26.5$
and $R-I<0.9$) is currently poorly known, we cannot give a definite
measure of the mass for these clusters.  If the mass of the clusters
can be measured by some other means, however, then one can measure the
mean redshift, and possibly the redshift distribution, of the
background galaxies.  We have done this (Clowe {\it et al.} 2000) and
find good agreement between the mean redshift we measure and the mean
redshift calculated from the Fontana {\it et al.} (1999) HDF-S
photometric redshift estimates, using the same magnitude and color
selection criteria. 

We have used two different methods to convert shears to convergences.
The first method is the KS93 inversion algorithm (Kaiser \&\ Squires
1993) which uses the fact that both the shear and the convergence are
combinations of second moments of the surface potential to transform,
in Fourier space, between the two.  This results in a two-dimensional
image of the convergence, and thus the surface density, of the
clusters but is limited in that it can only determine the convergence
to an unknown additive constant.  The second method, aperture
densitometry, measures the
circularly-averaged radial profile of the convergence around an
arbitrarily chosen center minus the average convergence of a chosen
annular region, usually set at the edges of the image.  One can then
either assume the convergence in this outer region is zero and measure
a minimum mass at a given radius or fit the profile with an assumed
mass model and determine the average convergence in the annular region
for the model.

\begin{figure}
\plotone{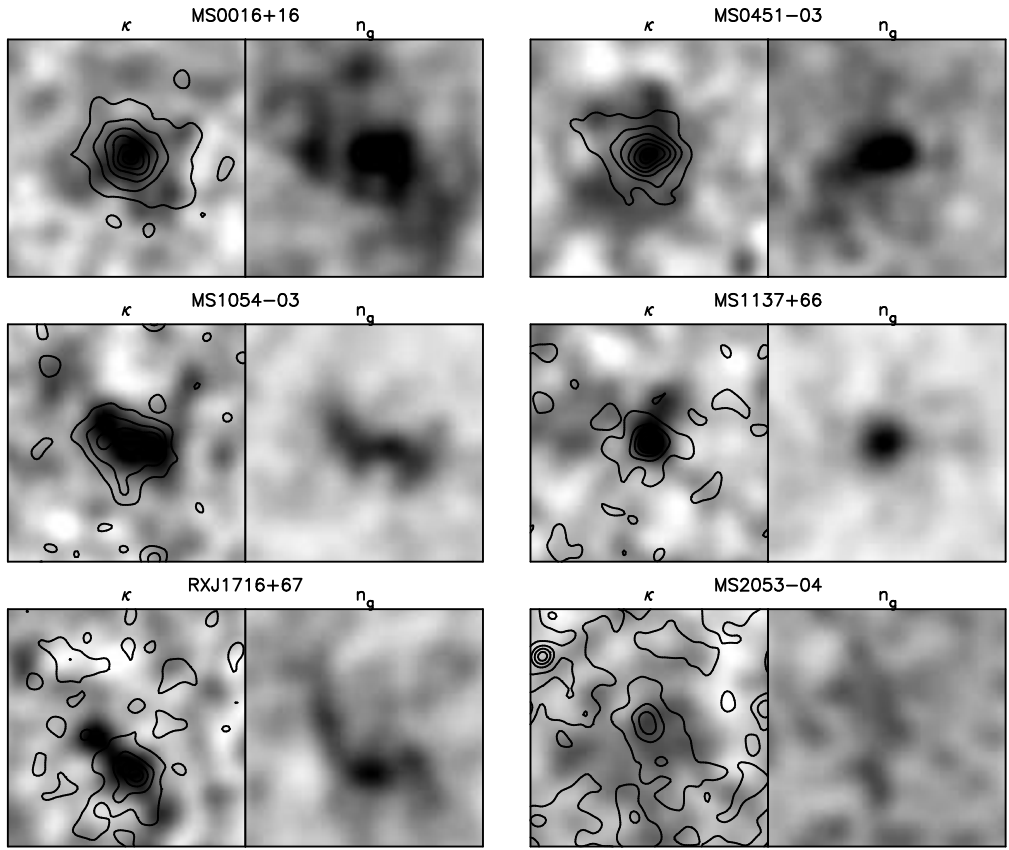}
\caption{Above are maps of the mass reconstruction (left), using the KS93
algorithm, and number density of cluster galaxies as selected by $R-I$
color (right) for all of the clusters in the survey.  Plotted in
contours overlayed on the mass reconstructions are X-ray images of the field from
the ROSAT PSPC.  All the images have been smoothed by the same scale,
and the greyscale is the same for every cluster.}
\end{figure}

\section{Discussion}

\begin{table}[t]
\caption{Cluster Sample}
{\scriptsize
\begin{tabular}{lccccccc}
\tableline
Cluster & z & $T_X$ & M (500 kpc) & M/L (500 kpc) & SIS & NFW & F-test\cr
&& keV & $h^{-1} 10^{14} M_\odot$ & $h M_\odot /L_\odot$ & $\chi ^2/(n-1)$ & $\chi
^2/(n-2)$ & \% \cr
\tableline
MS0016+16 & 0.547 & 8.4 & 4.1 & 260 & 1.08&0.93&85\cr
MS0451-03 & 0.550 & 10.4 & 7.9 & 480 & 1.61&1.32&97\cr
MS1054-03 & 0.826 & 12.3 & 12.7 & 590 & 1.33&0.94&99.5\cr
MS1137+67 & 0.783 & 5.7 & 6.1 & 670 & 1.13&1.16&-\cr
MS2053-04 & 0.586 & 8.1 & 3.0 & 360 & 0.99&1.00&-\cr
RX1716+67 & 0.809 & 6.7 & 3.8 & 350 & 1.00&1.03&-\cr
\tableline
\tableline
\end{tabular}
}
\end{table}

The maps of the convergence are given in Figure 1, along with contour
overlays of the ROSAT X-ray emission and maps of the number density of
galaxies with colors similar to the brightest cluster galaxy.  As can
be seen, there is in general a very good agreement between the
features present in the galaxy number density distribution and the
mass reconstructions.  Most of these features are, however, at
moderate statistical significance in the mass reconstructions,
and their shapes have probably been
altered to a large degree by the noise in the reconstructions.  In
particular, three of the clusters have a secondary peak in both the
mass and galaxy surface densities.  For two of these clusters,
MS1054$-$03 and RXJ1716$+$67, spectra of the galaxies in the peaks
have shown that they are at the same redshift as the main cluster peak
(Tran {\it et al.} 1999; Gioia {\it et al.} 1999), while
insufficient numbers of spectra exist for MS2053$-$04 to determine if the
second peak is physically associated with the cluster.  

Using the aperture densitometry profiles, assuming the center of each
cluster is the position of the brightest cluster galaxy, we have
calculated the best fit singular isothermal sphere and NFW profiles
(Navarro, Frenck, \&\ White 1996).  This was done using a $\chi ^2$ fitting
algorithm and using the mean redshift of the 
HDF-S photometric redshift catalog for galaxies with the same
magnitude and color range as the selected background galaxies.  We
find that the isothermal sphere model is a good fit to the lower mass
clusters, but a poor fit to the higher mass clusters.  The NFW
profiles, however, provide a good fit to all the clusters, although
the best fit profiles do not follow the relationship between total
mass and concentration as given by the zero redshift relaxed N-body
clusters in Navarro, Frenck, \&\ White (1996).  Using an F-test, we
have determined that the difference in the reduced $\chi ^2$ for the
two fits in the most massive clusters is significant in the 2-3
$\sigma$ range.  Masses for the clusters measured at a 500 $h^{-1}$
kpc radius from the BCG are given in table 1, along with the
mass-to-light ratio at the same radius using the luminosity of
galaxies with colors similar to the BCG to calculate the cluster
luminosity.  

We are currently investigating statistics which can be used to
quantify the amount of substructure present in the mass
reconstructions.  One statistic which we have tried is measuring
the ellipticity of the mass peaks from their second moments of the
surface density.  To do this, however, we first must
assume a value for the additive constant for each reconstruction.  We
have chosen this value so that the mean convergence in an
annular region located 2/3rds of the distance from the center to the
nearest edge in the image is the same as the mean convergence in the
best fit NFW profile to each cluster.  We then measure the second
moments using a Gaussian weighting function with a FWHM of 100
$h^{-1}$ kpc.  Ellipticities measured in this manner are extremely
sensitive to the value of the added constant, and changing the
constant from the maximum to minimum value allowed within the
1$\sigma$ NFW profiles from the aperture densitometry fits would often
change the measured ellipticity by a factor of two.  

To determine
the significance of these ellipticities we performed Monte-Carlo simulations
in which background galaxies were randomly positioned in a field
with the same number density as seen in the images.  These galaxies
were then sheared and displaced appropriately for an isothermal sphere
of mass similar to that measured in the clusters.  A mass
reconstruction was then performed on the galaxy distribution and the
ellipticity of the central peak was measured.  As the isothermal
spheres have a zero ellipticity, any ellipticity measured is induced
by the noise in the reconstruction from the intrinsic ellipticity
distribution of the galaxies.  From this distribution, we determined
that the minimum ellipticities measured above for MS1054, RXJ1716, and
MS0451 are greater than 72\%, 88\%, and 92\% of the simulations.  The
other clusters have ellipticities consistent with those induced by
noise.

A second statistic we have investigated is that of the separation
between the centroid of the mass distribution in the reconstruction, 
measured from minimizing the first moment of the mass distribution,
and the position of the brightest cluster galaxy.  To determine the
significance of the separations we used the same Monte-Carlo
simulations as above and measured the separation in centroid of the
reconstructed mass distribution and that used when shearing the
galaxies.  From this we find that separation seen in RXJ1716 is larger
than 98\% of the simulations, but that none of the other clusters has a
significant separation.

\acknowledgments
We wish to thank Pat Henry, Harald Ebeling, Chris Mullis, and Megan
Donahue for sharing their X-ray data prior to publication.  This work
was supported by NSF Grant AST-9500515 and the
``Sonderforschungsbereich 375-95 f\"ur
Astro--Teil\-chen\-phy\-sik" der Deutschen For\-schungs\-ge\-mein\-schaft.

\end{document}